\documentclass[fleqn,10pt]{wlscirep}
\usepackage[utf8]{inputenc}
\usepackage[T1]{fontenc}
\usepackage{amsmath}
\usepackage{braket}
\newcommand{\B}{\begin{equation}}
\newcommand{\E}{\end{equation}}

\title{A discrete relativistic spacetime formalism for 1+1-QED with continuum limits}

\author[1,*,+]{Kevissen Sellapillay}
\author[2,+]{Pablo Arrighi}
\author[3,$\dagger$,+]{Giuseppe Di Molfetta}
\affil[1]{Aix-Marseille Université, CPT, Campus de Luminy, case 907, 13288 Marseille, France}
\affil[2]{Université Paris-Saclay, CNRS, Laboratoire de recherche en informatique, 91405, Orsay, France and IXXI, Lyon}
\affil[3]{Aix-Marseille Université, Université de Toulon, CNRS, LIS, 13288, Marseille, France}

\affil[*]{kevissen.sellapillay@univ-amu.fr}
\affil[$\dagger$]{giuseppe.dimolfetta@lis-lab.fr}
\affil[+]{these authors contributed equally to this work}

\keywords{Schwinger model, QED, Quantum Cellular Automata, Dirac equation}

\begin{abstract}
We build a quantum cellular automaton (QCA) which coincides with $1+1$ QED on its known continuum limits. It consists in a circuit of unitary gates driving the evolution of particles on a one dimensional lattice, and having them interact with the gauge field on the links. The particles are massive fermions, and the evolution is exactly $U(1)$ gauge-invariant. We show that, in the continuous-time discrete-space limit, the QCA converges to the Kogut-Susskind staggered version of $1+1$ QED. We also show that, in the continuous spacetime limit and in the free one particle sector, it converges to the Dirac equation---a strong indication that the model remains accurate in the relativistic regime.
\end{abstract}
\begin{document}

\maketitle

\section*{Introduction}
Quantum physical phenomena can always be modelled classically by means of matrices and vectors. But, as far as we know, the dimension of these vectors grows exponentially with the number of particles, making these models intractable for classical computers. To simulate quantum physical phenomena efficiently, it seems we have no choice but to harness the laws of quantum mechanics themselves, as Feynman first suggested \cite{feynman_simulating_1982}. Quantum simulation could be applied to better understand condensed matter problems \cite{lloyd_universal_1996}, simulate molecules, find ground states of Hamiltonians, or even simulate the dynamics of quantum field theories (QFT) \cite{banuls_simulating_2020, preskill_simulating_2019, jordan_quantum_nodate}. It is the latter application that motivates this paper.

Amongst QFT, gauge theories are of fundamental importance to Physics, as they capture the fundamental interactions. Some of them have been recast in discrete space. Lattice QCD \cite{knechtli_lattice_2017} is the most famous example as it is extensively used to obtain theoretical numerical values, to be compared against experimental values coming out of particle accelerators: this procedure is partly how physicists are searching for new physics. Simulation has therefore taken a central role in the scientific method of particle physics. But these techniques are computationally heavy: finding a way to simulate lattice gauge theories efficiently and accurately with a quantum simulation device would be a game changer. Lattice gauge theories are also key for condensed matter through their application in spin liquids, and for quantum error correction e.g. via Kitaev's toric code \cite{savary_quantum_2016, kitaev_fault-tolerant_2003-1}.

The $1+1$ QED, also known as the Schwinger model \cite{schwinger_gauge_1962}, is a good candidate for a first step towards the quantum simulation of the dynamics of a gauge theory. Indeed, it is based on the $U(1)$ gauge group just like $3+1$ QED. It captures many non trivial physical properties such as a mass gap, fermion confinement and chiral symmetry breaking \cite{melnikov_lattice_nodate}. It is exactly solvable in the massless limit \cite{schwinger_gauge_1962}. These features explain why it is often used as a testbed for new techniques and ideas.

The standard ways to quantum simulate QFT are fundamentally non-relativistic, as
they all begin by expressing the theory in continuous-time discrete-space Hamiltonian form, using Kogut-Susskind methods \cite{kogut_hamiltonian_1975,banks_strong-coupling_1976}. They then map the matter (fermions) and the gauge field (bosons) onto quantum systems on a lattice, whose interactions will mimic those of the target Hamiltonian \cite{banuls_simulating_2020}. Sometimes these interactions are implemented as discrete-time products of quantum gates, but even then these are obtained by approximating the target Hamiltonian via the Trotter formula, an approximation which remains valid only in the non-relativistic regime $\Delta_t \ll \Delta_x$. This approach was recently realized experimentally on an ion trap architecture \cite{martinez_real-time_2016-1}. Numerical techniques exist that come to complement the standard approach, based on tensors networks. Those use compact, approximate descriptions of quantum states \cite{orus_practical_2014-1} such as the Density Matrix Renormalization Group (DMRG) \cite{byrnes_density_2002,zapp_tensor_2017-2}, discarding unwanted information about the states as they evolve, so that their description remain of manageable size---whilst attempting to keep track of the interesting physical ingredients.


In order to achieve quantum simulation in the relativistic regime $\Delta_x\approx \Delta_t$ one must keep space and time on an equal footing, discretizing both at the same time. This is sometimes referred to as digital quantum simulation. Digital simulation has indeed been very successful at describing relativistic particles in different fields \cite{MolfettaDebbasch2014Curved}, but so far it has not been able to produce simulation scheme for interacting QFT in the sector of more than 2 particles. 

A proposal was made in \cite{arrighi_quantum_2020}. From a QCA simulating relativistic Dirac equation a $U(1)$ gauge-invariant model was obtained. The convergence to the Schwinger model was not shown however and in fact, using the method presented in this paper, it can be shown that the Hamiltonian in the discrete space continuous time limit is not the correct one. We present here the circuit that gives the correct Hamiltonian in this limit.

In \cite{di_molfetta_quantum_2019, manighalam2021continuous} a quantum walk (i.e. the one particle sector of a QCA) was proposed which unifies non-relativistic analog quantum simulation with relativistic digital quantum simulation. Just by imposing $\Delta_x=\Delta_t^{1-\alpha}$ and tuning the $\alpha$, the operator is found to have well defined limits lattice fermions both in continuous-time discrete-space, and the relativistic Dirac equation in the continuous spacetime limit---a property referred to as \textit{plasticity}.

The QCA presented in this paper is closely related to these last two models. It is again based upon a QCA that recovers the relativistic Dirac equation, extended to become natively discrete gauge-invariant as in \cite{arrighi_quantum_2020}. But this time, the QCA is plastic, allowing us to prove its continuum limit towards $1+1$ QED, in the regime where $1+1$ QED does have a limit, i.e. the non-relativistic regime. In other words, we recover the Hamiltonian of the Kogut-Susskind Schwinger model in the continuous-time discrete-space limit. In the continuous spacetime limit we show that the QCA yields the Dirac equation in the free one particle sector, allowing to make the bridge between the non-relativistic and the relativistic regimes. Altogether, the QCA coincides with $1+1$ QED on its mathematical continuum limits, whenever these are defined.

The natively discrete digital circuit for staggered Schwinger model we propose is not seen in the literature. The QCA is staggered which is not usual in the QCA formalism \cite{arrighi_overview_2019-2}. One may wonder whether the approach, beyond the quantum simulation application, could be used to reframe QFT. Indeed, the fact that the QCA is gauge-invariant by construction, contains explicit relativistic and non-relativistic limits, is expressible by means of path integrals \cite{dariano_path-sum_2017,bisio_scattering_2019}, suggests that Quantum Computing point of view upon QFT may bring both rigour and pedagogy to the table---reviving the line of thought initiated by Feynman with his checkerboard propagator for 1+1 Dirac equation \cite{feynman_quantum_2010}.

The paper is organized as follows. We first define the QCA model, that is the spacetime structure and the gates. Second we show the continuous-time and discrete-space limit towards the Kogut-Susskind version of the Schwinger model, by means of the Jordan-Wigner transformation from qubits to fermions. Third we show that in a continuous spacetime limit, we recover the Dirac equation for the free one particle sector. Finally, we prove that the model is gauge-invariant and conclude by giving some perspectives. 

\section*{Model} 

\subsection*{The Kogut-Susskind staggered version of the Schwinger model}

The Schwinger model \cite{schwinger_gauge_1962} is a $(1+1)$D model invariant under the $U(1)$ gauge group. It models spinless electrons and their antiparticles, positrons, propagating on a 1D lattice and interacting with a $U(1)$ gauge field. We briefly summarize it by giving its Hamiltonian, which can be written using a temporal gauge ($A_0(x)=0$, and $A(x)=A_1(x)$) as :

\begin{equation}
\label{SchwingerHam}
H=\int \!\mathrm{d}x \bigl(\psi^{\dagger}(x)\left[\left(i\partial_x+i g A(x)\right)\sigma_z+m\sigma_x\right]\psi(x)+\frac{1}{2}E^2(x)\bigr),
\end{equation}
where $E(x)$ is the electric field observable at $x$, and $A(x)$ is its conjugate momentum, meaning
\begin{equation}
    [A(x),E(y)]=i\delta(x-y).
\end{equation}
Here $\psi(x)=(\psi_1(x),\psi_2(x))^T$ is a two components fermion field satisfying
\begin{equation}
    \{\psi_{\alpha}(x),\psi^{\dagger}_{\beta}(y)\}=\delta_{\alpha\beta}\delta(x-y).
\end{equation}
\begin{figure}[h!]
	\begin{center}
  \includegraphics[trim = 0cm 0cm 0cm 0cm, clip, width=10 cm]{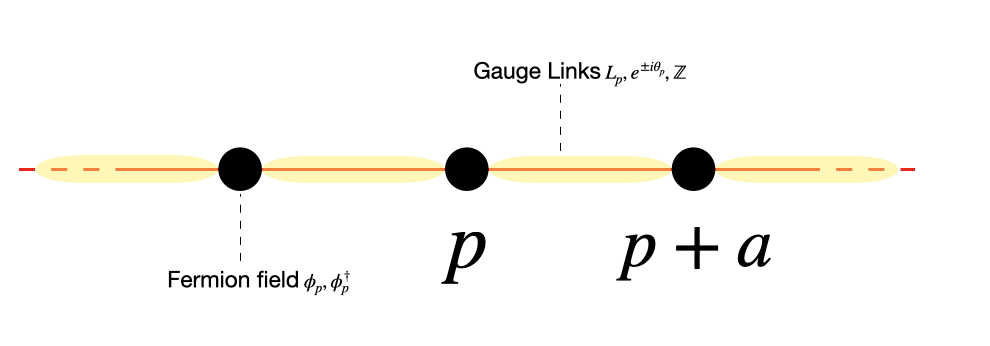}
  \caption{\label{fig:link}Kogut-Susskind version of the Schwinger model. The gauge (boson) field is represented by the links in red. These links are states $\ket{l}$ where $l$ takes value in $\mathbb{Z}$. The operators acting on them are $L_{p}$ and $e^{\pm i\theta_{p}}$. The matter (fermion) field $\phi_{p}$, $\phi^{\dagger}_{p}$ are on the nodes in black. Even (odd) sites correspond to upper (lower) component of a spinor field. $a$ is the lattice spacing. }
\end{center}

\end{figure}
We now describe the staggered Kogut-Susskind version of the Schwinger model. This Kogut-Susskind procedure consists in putting fermion fields on the nodes of an infinite 1D lattice and bosonic gauge fields on the links between them, as depicted in Fig.\ref{fig:link}. We can interpret occupied odd sites as electrons and unoccupied even sites as positrons, therefore particles and antiparticles are described by a single fermion field \cite{martinez_real-time_2016-1} (staggered picture). This procedure allows to give a continuous time, discrete space formulation of a continuous spacetime model and is also a way to partly resolve the fermion doubling problem. 

The Hamiltonian of this staggered version reads \cite{kogut_hamiltonian_1975,banks_strong-coupling_1976}:
\begin{equation}
\centering
\boxed{
H_{S}=\frac{i}{2a}\sum\limits_{p}(\phi^{\dagger}_{p+1}e^{-i\theta_{p}}\phi_{p}-h.c.)+m\sum\limits_{p}(-1)^{p}\phi^{\dagger}_{p}\phi_{p} +\frac{ag^{2}}{2}\sum\limits_{p}L^{2}_{p}},
\end{equation}
where $a$ is the lattice spacing, $g$ the strength of the interaction (the charge of the particles), $m$ the mass and $\phi$ the fermion field. To construct the spinor $\Psi_{\tilde{p}}$ from the fermion fields, we group fields in pairs $(p,p+1)$, where $p$ is even.
In between the fermion fields at $p$ and $p+1$, there is a gauge field link which takes values in $\mathbb{Z}$. The operators $e^{\pm i\theta_{p}}$ raises or lowers the value of this link $[p,p+1]$ such that : 
\begin{equation}
e^{\pm i\theta_{p}}\ket{l}_{p}=\ket{l\pm1}_{p}
\end{equation}
and the electric field is given by $E_{p}=gL_{p}$ where 
\begin{equation}
L_{p}\ket{l}_{p}=l\ket{l}_{p}.
\end{equation}

\subsection*{A quantum cellular automaton for 1+1 QED}

\begin{figure}[h!]
	\begin{center}
  \includegraphics[trim = 0cm 0cm 0cm 0cm, clip, width=10 cm]{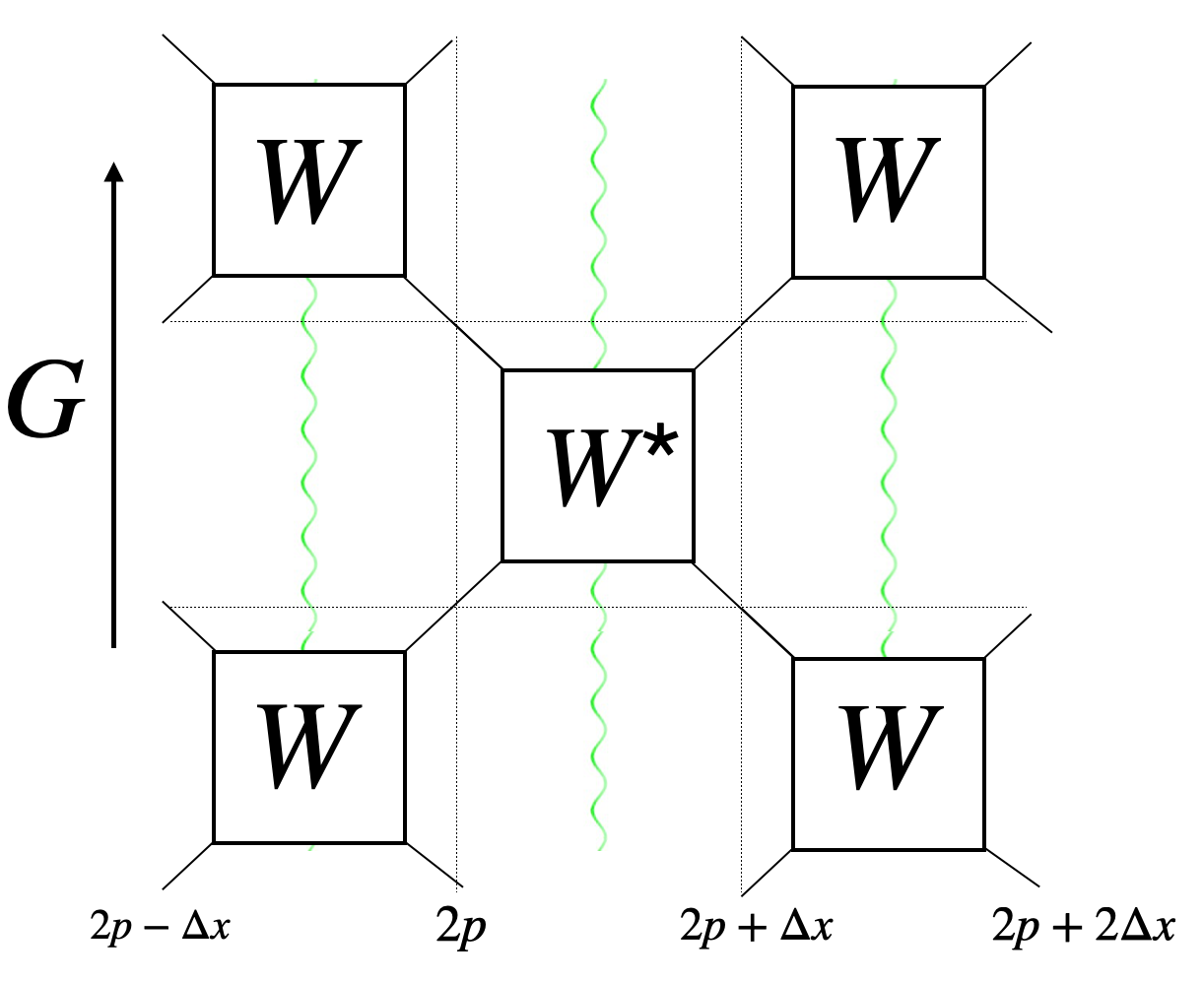}
  \caption{\label{fig:QCA}QCA structure. One application of $G$, the evolution operator, corresponds to two rows, first a row of $W$ and then of $W^{*}$ gates. Black wires represent the fermionic fields, green wires represent the gauge field. The qubits are separated in space by $\Delta x$ and the gates are separated in time by $\Delta t$.}
\end{center}

\end{figure}

The model we propose consists in having one qubit per site $p$ separated by $\Delta x$ and gauge fields located on the links between each sites, at half step, modeled by states taking values in $\mathbb{Z}$ \cite{arrighi_quantum_2020}. This gauge field could be experimentally represented by qudits or harmonic oscillators. We choose the evolution operator of the QCA to be : 
\begin{equation}
\label{eqn:evolution}
G=\underset{p' \text{ even}}{\otimes}W^{*}_{p'}e^{-\frac{i}{2}\Delta x\Delta tg^{2}L^{2}}\underset{p\text{ odd}}{\otimes}W_{p}e^{-\frac{i}{2}\Delta x\Delta tg^{2}L^{2}}.
\end{equation}
where  
\begin{equation}
\label{eqn:Wmat} W_{p}=\begin{pmatrix}
I && 0 && 0 && 0 \\
0 && e^{-i\zeta}\sin\theta I && \cos\theta V_{p+\frac{1}{2}} && 0 \\
0 && -\cos\theta V^{\dagger}_{p+\frac{1}{2}} && e^{i\zeta}\sin\theta I  && 0 \\
0 && 0 && 0 && I
\end{pmatrix}. 
\end{equation}
corresponds to the fermions dynamics and the exponential applied before each $W_{p}$ codes the interaction with the bosons field at the same position. In particular, as depicted in Fig. \ref{fig:QCA}, the gate $W_{p}$ and the gate $W^*_{p}$ are located in the space-time grid in between the qubit at position $(p,p+1)$ respectively for \textit{odd} and \textit{even} $p$. Each gate $W$ and its conjugate acts on the local gauge field. Moreover, in order to avoid the fermion doubling problem we choose to work with a staggered QCA, in which occupied odd sites are interpreted as electrons and unoccupied even sites as positrons. This is consistent with the same interpretation that Kogut and Susskind gave of the Schwinger model. 
From a dynamical point of view, the matrix $W_{p}$ can be interpreted as follows~: diagonal terms correspond to staying on the same site, eventually picking up a phase related to the mass. The non diagonal part correspond to hopping terms~: a right-moving $\ket{1}$ will decrease the gauge field it passes through, a left-moving one will increase it. The gauge field operators read: 
\begin{equation}
V_{p+\frac{1}{2}}\ket{l}_{p+\frac{1}{2}}=\ket{l-1}_{p+\frac{1}{2}},
\end{equation}
and \B V^{\dagger}_{p+\frac{1}{2}}\ket{l}_{p+\frac{1}{2}}=\ket{l+1}_{p+\frac{1}{2}}.\E
The identities $\mathbb{I}$ means that the gauge link is left invariant. The operator $L$ in the exponentials of Eq. \eqref{eqn:evolution} acts on the states of the gauge field $\ket{l}_{p+\frac{1}{2}}$ in the following way : 
\B L\ket{l}_{p+\frac{1}{2}}=l\ket{l}_{p+\frac{1}{2}},\E for $l\in\mathbb{Z}$.\\

Now, in order to map the above automaton with the Schwinger model we move to the second quantization formalism. Let us start with the gate $W_{p}$ (\ref{eqn:Wmat}), which acts on a pair of qubits $(p,p+1)$ and a gauge link $[p,p+1]$. It is useful to introduce $\tilde{W}_{p}=I_{<p}\otimes W_p \otimes I_{>p}$. 
Each $W_p$ can be written in terms of single qubit operators $E_{ij}=\ket{i}\bra{j}$. Using the order  $(0_{p} 0_{p+1},0_{p} 1_{p+1},1_{p} 0_{p+1},1_{p} 1_{p+1})$ and omitting identities on the gauge link for clarity, we have : 
 \begin{align} \begin{split} W_{p}=(E_{00})_{p} \otimes (E_{00})_{p+1} + e^{-i\zeta}\sin\theta (E_{00})_{p} \otimes (E_{11})_{p+1}+\\
 e^{i\zeta}\sin\theta (E_{11})_{p} \otimes (E_{00})_{p+1} +\cos\theta V_{p+\frac{1}{2}}(E_{01})_{p} \otimes (E_{10})_{p+1}\\
 -\cos\theta V^{\dagger}_{p+\frac{1}{2}}(E_{10})_{p} \otimes ( E_{01})_{p+1} + (E_{11})_{p} \otimes ( E_{11})_{p+1},\end{split} \end{align} 
Thus, we ought to transform qubit operators to fermionic operators, via the the standard Jordan-Wigner transformations, and finally introduce the following annihilation and creation operators :
\begin{align} \begin{split}&\phi_{p}=I_{<p}\otimes (E_{01})_{p}\otimes (\sigma_{z})_{p+1}\otimes (\sigma_{z})_{p+2}...\\&
\phi^{\dagger}_{p}=I_{<p}\otimes(E_{10})_{p}\otimes(\sigma_{z})_{p+1}\otimes (\sigma_{z})_{p+2}...,
\end{split}\end{align}
with $\{ \phi_{p}, \phi^{\dagger}_{p'} \}=\delta_{p,p'}I$ and $\{ \phi_{p}, \phi_{p'} \}=0$. Notice that the $\sigma_{z}$ to the right are specifically there to ensure the correct anti-commutation relation (Supplementary Material 1.), together they form the string operator of the Jordan-Wigner transformation. 
Looking at these operators, we see that creating or annihilating a fermion on a qubit lattice is very much a non-local operation and we could be worried that the resulting dynamics using such operators would be non-local and unphysical. However, the terms involving these operators in the Hamiltonian are quadratic, therefore they become perfectly local and physical. This can already be seen at the level of the gates written in this formalism. In fact, for $\tilde{W}_{p}$ we have :
\begin{align} \begin{split} \tilde{W}_{p}&=\phi_{p}\phi^{\dagger}_{p}\phi_{p+1}\phi^{\dagger}_{p+1} +e^{-i\zeta}\sin\theta \phi_{p}\phi^{\dagger}_{p}\phi^{\dagger}_{p+1}\phi_{p+1}+e^{i\zeta}\sin\theta \phi^{\dagger}_{p}\phi_{p}\phi_{p+1}\phi^{\dagger}_{p+1} \\&-\cos\theta V_{p+\frac{1}{2}} \phi_{p}\phi^{\dagger}_{p+1}-\cos\theta V^{\dagger}_{p+\frac{1}{2}} \phi^{\dagger}_{p}\phi_{p+1}+\phi^{\dagger}_{p}\phi_{p}\phi^{\dagger}_{p+1}\phi_{p+1}.
\end{split} \end{align} 
Putting together these transformed gates, the global evolution reads:
\begin{align} \begin{split}
G&
=\prod\limits_{p'  \text{ even}}\tilde{W}^{*}_{p'}e^{-\frac{i}{2}\Delta x\Delta tg^{2}L^{2}}\prod\limits_{p \text{ odd}}\tilde{W}_{p}e^{-\frac{i}{2}\Delta x\Delta tg^{2}L^{2}}.
\end{split} \end{align} 

\section*{Methods}
\subsection*{Continuous limits}

In order to prove that the above staggered QCA, reformulated in terms of fermionic operators, converges to the Schwinger Hamiltonian, we introduce the following parametrization :
\begin{eqnarray}
\Delta t &=& \epsilon \nonumber \\
\Delta x &=& \epsilon^{1-\alpha} \nonumber \\
\kappa&=&\epsilon^{\alpha}\\
\theta&=&\arccos(c\kappa) \nonumber \\
\zeta&=&m\frac{(-1)^{\kappa}\epsilon}{\sin(\theta)} \nonumber 
\label{eqn:param}
\end{eqnarray}
where the case $\alpha = 1$ and $\alpha = 0$ correspond respectively to the continuous time and discrete space limit and the continuous spacetime limit. \\

In the non relativistic limit, the evolution is continuous in time and discrete in space. The dynamics is driven by a Hamiltonian, which is recovered looking at the first order of the global evolution operator of the QCA, as follows: 
\begin{equation}
\label{eqn:dyna} 
G= e^{-2i H_{QCA} \Delta t}\simeq 1-2i \Delta t H_{QCA}.
\end{equation}
We then take the limit $\epsilon \rightarrow 0$ on $G$, using the parametrisation \eqref{eqn:param}, for $\alpha=1$. Finally we get :
\begin{align} \begin{split} \tilde{W}_{p}&\simeq \phi_{p}\phi^{\dagger}_{p}\phi_{p+1}\phi^{\dagger}_{p+1} +(1-im\epsilon) \phi_{p}\phi^{\dagger}_{p}\phi^{\dagger}_{p+1}\phi_{p+1}+(1+im\epsilon) \phi^{\dagger}_{p}\phi_{p}\phi_{p+1}\phi^{\dagger}_{p+1} \\&-\epsilon V_{p+\frac{1}{2}} \phi_{p}\phi^{\dagger}_{p+1}-\epsilon V^{\dagger}_{p+\frac{1}{2}} \phi^{\dagger}_{p}\phi_{p+1}+\phi^{\dagger}_{p}\phi_{p}\phi^{\dagger}_{p+1}\phi_{p+1}\\&
=1 + \epsilon \Big[\phi^{\dagger}_{p+1}\phi_{p}V_{p+\frac{1}{2}} -  V^{\dagger}_{p+\frac{1}{2}} \phi^{\dagger}_{p}\phi_{p+1}+im(\phi^{\dagger}_{p}\phi_{p}\phi_{p+1}\phi^{\dagger}_{p+1}-\phi_{p}\phi^{\dagger}_{p}\phi^{\dagger}_{p+1}\phi_{p+1})\Big].
\end{split} \end{align} 
We can rewrite the mass term :
 \begin{align*} \begin{split}& \phi^{\dagger}_{p}\phi_{p}\phi_{p+1}\phi^{\dagger}_{p+1}-\phi_{p}\phi^{\dagger}_{p}\phi^{\dagger}_{p+1}\phi_{p+1} = \phi^{\dagger}_{p}\phi_{p}\phi_{p+1}\phi^{\dagger}_{p+1}-(1-\phi^{\dagger}_{p}\phi_{p})\phi^{\dagger}_{p+1}\phi_{p+1}\\&
 \phi^{\dagger}_{p}\phi_{p}(\phi_{p+1}\phi^{\dagger}_{p+1}+\phi^{\dagger}_{p+1}\phi_{p+1})-\phi^{\dagger}_{p+1}\phi_{p+1}=\phi^{\dagger}_{p}\phi_{p}-\phi^{\dagger}_{p+1}\phi_{p+1}.
\end{split} \end{align*} 

Finally we have

\B \tilde{W}_{p}\simeq 1 + \epsilon \Big[\phi^{\dagger}_{p+1}\phi_{p}V_{p+\frac{1}{2}} -  V^{\dagger}_{p+\frac{1}{2}} \phi^{\dagger}_{p}\phi_{p+1}+im(\phi^{\dagger}_{p}\phi_{p}-\phi^{\dagger}_{p+1}\phi_{p+1})\Big],
\E 

and the interaction with the gauge field can be developped as 

\B e^{-\frac{i}{2}\Delta x\Delta tg^{2}L^{2}} \simeq  1-\frac{i}{2}\epsilon g^{2} \sum\limits_{i} L_{i}^{2} .\E

Combining everything, we get for $G$ : 
\begin{align} \begin{split}
G&=\prod\limits_{p'  \text{ even}}\tilde{W}^{*}_{p'}e^{-\frac{i}{2}\Delta x\Delta tg^{2}L^{2}}\prod\limits_{p \text{ odd}}\tilde{W}_{p}e^{-\frac{i}{2}\Delta x\Delta tg^{2}L^{2}}\\&
=\prod\limits_{p'}(1+\epsilon[\phi^{\dagger}_{2p'+1}\phi_{2p'}V_{2p'+\frac{1}{2}} -  \phi^{\dagger}_{2p'}\phi_{2p'+1}V^{\dagger}_{2p'+\frac{1}{2}} -im(\phi^{\dagger}_{2p'}\phi_{2p'}-\phi^{\dagger}_{2p'+1}\phi_{2p'+1})])\\&
(1-\frac{i}{2}\epsilon g^{2} \sum\limits_{i}L_{i}^{2})
\prod\limits_{p}(1+\epsilon[ \phi^{\dagger}_{2p+2}\phi_{2p+1}V_{2p+\frac{3}{2}} -  \phi^{\dagger}_{2p+1}\phi_{2p+2}V^{\dagger}_{2p+\frac{3}{2}}\\&+im(\phi^{\dagger}_{2p+1}\phi_{2p+1}-\phi^{\dagger}_{2p+2}\phi_{2p+2})])(1-\frac{i}{2}\epsilon g^{2} \sum\limits_{i}L_{i}^{2}).
\end{split} \end{align}

A straightforward calculation leads us to the leading order of the series :
\begin{align} \begin{split}
\label{eqn:Geq}
G&\simeq 1+\epsilon\sum\limits_{p}[\phi^{\dagger}_{2p+1}\phi_{2p}V_{2p+\frac{1}{2}} -  \phi^{\dagger}_{2p}\phi_{2p+1}V^{\dagger}_{2p+\frac{1}{2}} + 
\phi^{\dagger}_{2p+2}\phi_{2p+1}V_{2p+\frac{3}{2}} -  \phi^{\dagger}_{2p+1}\phi_{2p+2}V^{\dagger}_{2p+\frac{3}{2}}
 \\&-i g^{2} L_{p}^{2}
-im(\phi^{\dagger}_{2p}\phi_{2p}-2\phi^{\dagger}_{2p+1}\phi_{2p+1}+\phi^{\dagger}_{2p+2}\phi_{2p+2})].
\end{split} \end{align} 
Notice that the mass term can be rewritten as : $-2im(\phi^{\dagger}_{2p}\phi_{2p}-\phi^{\dagger}_{2p+1}\phi_{2p+1})$
Moreover the four hopping terms simplify because each pair of two terms is separated by one step : 
\B\sum\limits_{p}\phi^{\dagger}_{p+1}\phi_{p}V_{p+\frac{1}{2}}-h.c.\E
Finally, we identify $V_{p+\frac{1}{2}}$ to $e^{-i\theta_{p}}$ and $V^{\dagger}_{p+\frac{1}{2}}$ to $e^{i\theta_{p}}$. Identifying the Hamiltonian in (\ref{eqn:Geq}) using  $G \simeq 1-2i \epsilon H_{QCA} $, we find the Hamiltonian of the QCA to be :
\begin{equation}
\boxed{H_{QCA}=\sum\limits_{p}[\frac{i}{2}(\phi^{\dagger}_{p+1}\phi_{p}e^{-i\theta_{p}}-h.c)
+m(-1)^{p}\phi^{\dagger}_{p}\phi_{p}
+\frac{g^{2}}{2} L_{p}^{2}]}.
\end{equation} 
The above one coincides with the Kogut-Susskind Hamiltonian of the Schwinger model $H_{S}$ with $a=1$. We have thus identified a QCA-based quantum simulator for a QED toy model, namely a theory for both spinless electrons and positrons and their interaction with a dynamical gauge field. \\


We ought to be sure that in the relativistic limit our simulator reproduces the right dynamics. Here, we give a proof of that in the simplest scenario, the non-interacting case. Starting from the the one particle sector of the staggered QCA, we take the $\alpha=0$ relativistic limit, namely for $\Delta t =\Delta x = \epsilon\rightarrow 0$. (\ref{eqn:param}).
The local gate which drives the automaton simplifies as follow
\B W'_{p}=\begin{pmatrix}
1 && 0 && 0 && 0 \\
0 && e^{-i\zeta}\sin\theta && \cos\theta && 0 \\
0 && -\cos\theta  && e^{i\zeta}\sin\theta  && 0 \\
0 && 0 && 0 && 1
\end{pmatrix},\E
and the global evolution operator reads:
\begin{align} \begin{split}
G=\underset{p' \text{ even}}{\otimes}W^{*}_{p'}\underset{p\text{ odd}}{\otimes}W_{p},
\end{split}\end{align}
Let's start with a general 1 particle state : $\ket{\Psi(t)}=\sum\limits_{x}\psi(t,x)\ket{1}_{x}$ and separate the one particle state into pairs of even and odd sites :  
\begin{align} \begin{split}\label{eqn:stag}
\ket{\Psi(t)}=\sum\limits_{x \text{ even} }\psi^l(t,x)\ket{1}_{x} + \psi^r(t,x+\epsilon)\ket{1}_{x+\epsilon},
\end{split}\end{align}
After one time step evolution of the automaton, the recurrence relations on the amplitudes $\mathbf{s}\equiv(\psi^l,\psi^r)$ which governs the single fermion, reads:
\begin{align} 
\mathbf{s}(t+2\epsilon,x)=&\begin{pmatrix}
0 & -\cos\theta e^{-i\zeta}\sin\theta \\
0 & \cos^{2}\theta
\end{pmatrix} \mathbf{s}(t,x-2\epsilon)
+\begin{pmatrix}
e^{-2i\zeta}\sin^{2}\theta & \cos\theta e^{i\zeta}\sin\theta \\
-\cos\theta e^{-i\zeta} \sin\theta & e^{2i\zeta}\sin^{2}\theta
\end{pmatrix}\mathbf{s}(t,x)\\&
+\begin{pmatrix}
\cos^{2}\theta & 0\\
\cos\theta e^{i\zeta}\sin\theta & 0
\end{pmatrix}\mathbf{s}(t,x+2\epsilon).
\end{align}

Taking the limit $\epsilon\rightarrow 0$, we find the following differential equation :

\B
\partial_{t}\mathbf{s}(t,x)=P \partial_{x}\mathbf{s}(t,x)+Q\mathbf{s}(t,x).
\label{eq:Dirac1}
\E
where the operators $P$ and $Q$ are represented in the computational basis  : 
\begin{equation}
P = \begin{pmatrix}
c^{2} & c \sqrt{1-c^{2}}\\
c\sqrt{1-c^{2}} & -c^{2}
\end{pmatrix}
\end{equation}
and 
\begin{equation}
Q = \begin{pmatrix}
im\sqrt{1-c^{2}} & -cim \\
-cim & -im\sqrt{1-c^{2}}
\end{pmatrix}.
\end{equation}
The operator $P$ is self-adoint and its eigenvalues are $\pm c$. Two eigenvectors associated to these eigenvalues are :
\begin{equation}
\begin{split}
    b_- = -\sqrt{\frac{1-c}{2}}b_0+\sqrt{\frac{1+c}{2}}b_1 \hspace{1cm}  b_+ =\sqrt{\frac{1+c}{2}} b_0 + \sqrt{\frac{1-c}{2}}b_1
    \end{split}
\end{equation}
The family $(b_-,b_+)$ forms an orthonormal basis of the two dimensional spin Hilbert space. Let us now rewrite equation \eqref{eq:Dirac1} in this new orthonormal basis. A straightforward computation leads to:


\B\label{eqn:DIRAC}
i\gamma^0\partial_{0}\mathbf{\tilde s}(t,x) + i \gamma^1\partial_{1}\mathbf{\tilde s}(t,x) - m\mathbf{\tilde s}(t,x) = 0,
\E
where $\gamma^0 = \sigma_{x}$, $\partial_{0} = \partial_t$, $\gamma^1 = \sigma_{x}\sigma_{z}$ and $\partial_{0} = c \partial_x$.\\

\subsection*{Remark}

Let us shortly discuss the last term of operator $W(p)$. In general, when two creation operators get exchanged, a minus sign is produced so as to respect the anti-commutation of their creation operators. The abrupt minus appearance of minus signs makes it harder to compute quantities about fermions---an issue which is sometimes referred to as the `sign problem'. Here, when $\Delta t \approx \Delta x$, we must indeed put a $-1$ in the bottom right coefficient of the gate, since two creation operators are crossing during the lapse of one $W$ gate, as was shown in detail \cite{arrighi_quantum_2020}. However, when $\Delta t \ll \Delta x$ we should not. Physically, this is because the two fermions now hardly have time to cross in the lapse of one gate. Mathematically, this shows through the fact that placing a $-1$ at this position forbids the development of the gate around identity, ruining any effort to obtain a non-relativistic continuous-time discrete-space limit towards the Kogut-Susskind Hamiltonian, or any other Hamiltonian for that matter.
 To get the best of both worlds, we use scaling factor $e^{i(\frac{\Delta t}{\Delta x})^{2}\pi}=e^{i\epsilon^{2\alpha}\pi}$, making the coefficient go to $1$ in the non-relativistic parametrization ($\alpha=1$) and to $-1$ in the relativistic parametrization ($\alpha=0$). This is compatible with unitarity, plasticity, and fermionic computation.
\begin{equation}W''_{p}=\begin{pmatrix}
1 && 0 && 0 && 0 \\
0 && \sin\theta  && -\cos\theta V_{p+\frac{1}{2}} && 0 \\
0 && \cos\theta V^{\dagger}_{p+\frac{1}{2}} && \sin\theta  && 0 \\
0 && 0 && 0 && e^{i\epsilon^{2\alpha}\pi}
\end{pmatrix}.\end{equation}
It is the choice that yields the Kogut-Susskind Hamiltonian in the non-relativistic regime. In the relativistic regime, however, $\alpha=0$ and so at order 0, we see that : 
\begin{align} \begin{split}  \tilde{W}_{p}& \simeq\phi_{p}\phi^{\dagger}_{p}\phi_{p+1}\phi^{\dagger}_{p+1} +\phi_{p}\phi^{\dagger}_{p}\phi^{\dagger}_{p+1}\phi_{p+1}+\phi^{\dagger}_{p}\phi_{p}\phi_{p+1}\phi^{\dagger}_{p+1} -\phi^{\dagger}_{p}\phi_{p}\phi^{\dagger}_{p+1}\phi_{p+1}\\&
=1-2\phi^{\dagger}_{p}\phi_{p}\phi^{\dagger}_{p+1}\phi_{p+1}.
\end{split} \end{align} 
Since this is a control-Z on the $2$ qubits, the order $0$ of $G$ cannot be the identity. As expected, we cannot recover a many-body interacting Hamiltonian in this regime.


\section*{The staggered QCA is gauge invariant}

\begin{figure}[h!]
	\begin{center}
  \includegraphics[trim = 0cm 0cm 0cm 0cm, clip, width=10 cm]{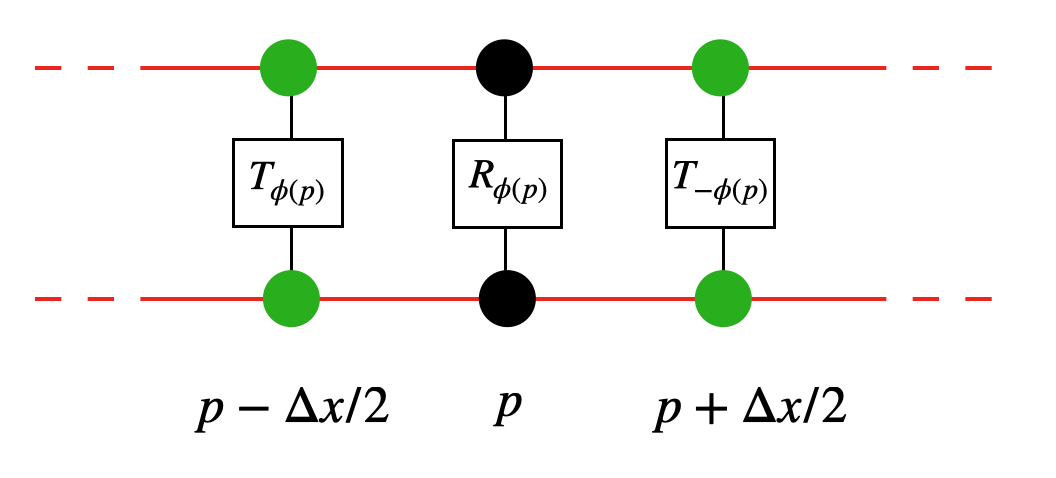}
  \caption{\label{fig:gauge}The gauge transformation. Black dots are the fermions sites. Green dots in between sites represent the gauge field values. At a single site $p$, the gauge transformation will apply a gate $R_{\varphi(p)}$  that gives a phase when the qubit there is in state $\ket{1}$ and apply gates $T_{\varphi(p)}$ on the left and right gauge field points that produce phases according to the values of the gauge field. This transformation is applied on every sites, therefore a single gauge field is acted upon twice.}
\end{center}
\end{figure}

In this section, we show that our QCA-based quantum simulator is gauge invariant at finite scale. We define the gauge transformation in Fig.\ref{fig:gauge}, where the gates are defined as follows :

\B
\label{eqn:defgauge}
P_{\varphi}=\underset{p}{\otimes}P_{\varphi(p)}=\underset{p}{\otimes}\Big( T_{\varphi(p)}\otimes R_{\varphi(p)}  \otimes T_{-\varphi(p)}\Big),
\E
where $ R_{\varphi(p)} $ acts on the qubit at site $p$ such that :
\begin{align}
R_{\varphi(p)} : &\ket{0} \rightarrow \ket{0}\\&
\ket{1}\rightarrow e^{i\varphi(p)}\ket{1},
\end{align}
and $T_{\varphi(p)}$ acts on the gauge field states such that :
\B
T_{\varphi(p)}\ket{l}=e^{il\varphi(p)}\ket{l}.
\E
From definition (\ref{eqn:defgauge}), we see that a gauge field at a given site $p+\frac{1}{2}$ will be acted upon twice, once by $-\varphi(p)$ and once by $+\varphi(p+1)$. Is our QCA invariant under the above gauge transformation, or namely:
\B
\label{eqn:gaugeTF}
P_{\varphi}G\stackrel{?}{=}GP_{\varphi}.
\E

Equation (\ref{eqn:gaugeTF}) means that whatever the field $\varphi(p)$ is, the evolution operator should give the same result if we apply it after the gauge transformation or before the gauge transformation, in other words, the dynamics should not be changed by this gauge field, the latter should have no physical consequences. However if we impose this $U(1)$ local phase, invariance is only possible if we introduce some interaction between qubits and the gauge fields , which is done through the $V$ and $V^{\dagger}$ operators, that will compensate the missing pieces of phase appearing during the particle movement, as done in \cite{arrighi_quantum_2020}. 
To show the gauge invariance (\ref{eqn:gaugeTF}), we start from a ket $\ket{1}$ at a given site (odd or even), with only gauge field values given in the rest of the space. We compute explicitly the results after applying first $P_{\varphi}$ then $G$ and the other way around, and show that the result are exactly the same, therefore the gauge transformation $P_{\varphi}$ has no physical consequences on the dynamics $G$. The model is thus gauge invariant (Supplementary Material 2.).

\section*{Conclusion}

We have described a quantum cellular automaton (QCA) that simulates $1+1$ QED. It consists in a lattice of qubits encoding whether a fermion is present at a given site. These interact with the gauge field that lives on the links between those sites. The QCA was shown to coincide with the Kogut-Susskind version of the Schwinger model in the continuous-time discrete-space limit, and with the Dirac equation in the continuous spacetime limit in the one-particle sector. We go from one limit to the other just by imposing $\Delta_x=\Delta_t^{1-\alpha}$ and tuning the $\alpha$. 

We still cannot ascertain the QCA recovers the Schwinger model in the interactive regime of the continuous spacetime limit, but then again it is not even clear that the lattice QFT has such a limit in the first place. The QCA coincides with the Schwinger model wherever it has a known, mathematically defined limit. 

It coincides in the story it tells; of fermions propagating relativistically updating the gauge field, which in turn simply triggers a phase---thereby turning on the interaction.  

Finally, it coincides in terms of its construction: the QCA retains the fundamental $U(1)$ gauge-invariance of QED, even in discrete spacetime. This gauge-invariance construction was originally proposed in \cite{arrighi_quantum_2020} for QCA, and in \cite{arrighi2018gauge} for Reversible CA. Both drew inspirations from gauge-invariant Quantum Walks \cite{di2012discrete, MolfettaDebbasch2014Curved, marquez2018electromagnetic}. The question of extending these constructions to non-abelian gauge theories has been treated for Quantum Walks \cite{arnault_quantum_2016-1,di2016quantum} and for Reversible CA \cite{ArrighiNonAbelianGauge}. Lifting this to obtain $U(N)$ gauge-invariant QCA is no doubt one of the next steps lying ahead towards digital quantum simulation schemes for Yang Mills theories, such as QCD. An observation was made in \cite{farrelly_discretizing_2020} that the path integral of such a theory can be written as a real-time transfer operator which is itself a finite-depth local quantum circuit.

Another obvious next step is the extension to $2+1$ and $3+1$ dimensions. Of particular concern is the fact that we relied upon the Jordan Wigner mapping to encode fermions into qubits. This transformation is known to suffer locality issues \cite{huerta_bose-fermi_1993, dariano_fermionic_2014, dariano_feynman_2014}, which did not affect us because our terms were quadratic, cancelling out all non-local effects. In further dimensions we may not be so lucky however, and will have to rely on alternative transformations \cite{farrelly_insights_2017, brun_quantum_2020, mlodinow_fermionic_2020}. A possible route of investigation could be to use formulations of QED in 2+1D or 3+1 dimensions where the Hilbert space dimension is reduced by either approximating the gauge group \cite{haase_resource_2021} or by going in a rotating frame \cite{bender_gauge_2020} which decouples the matter to the gauge field and keeps only local constraints on the latter. 

The problem of preparing the ground state of such QCA is puzzling as was pointed out in \cite{arrighi_quantum_2020}. We do not solve the problem but by recovering a Hamiltonian in the continuous-time discrete-space limit, we make the problem well-defined. 

QCA are closely related to path integrals \cite{dariano_path-sum_2017}. Lately a formalism was developed for dealing with interactions in a perturbative manner \cite{bisio_scattering_2019}, within the QCA framework---which would be an interesting application here. Another vast topic for exploration is trying to understand the link between renormalization theory, and the way the parameters of the unitary gates must be made to vary with the lattice as we take our limits. 

The plastic Quantum Walk \cite{di_molfetta_quantum_2019} upon which this plastic QCA is built is deformable to the point that a curved spacetime limit can been obtained just by putting a spacetime dependence in the parameter $c$. A fair question to ask is whether allowing for the same spacetime dependence here, would yield the Schwinger model on a curved background \cite{parker_quantum_2009}.

\section*{Acknowledgements}
We benefited from discussions with C\'edric B\'eny, Nathana\"el \'Eon, Terry Farrelly, Alberto Verga. This publication was made possible through the support of the ID\# 61466 grant from the John Templeton Foundation, as part of the “The Quantum Information Structure of Spacetime (QISS)” Project (qiss.fr). The opinions expressed in this publication are those of the author(s) and do not necessarily reflect the views of the John Templeton Foundation.

\bibliography{sample}
\end{document}